\documentclass[11pt]{article}
\usepackage{graphicx}
\usepackage{amsmath,amssymb}

\begin{document}

\title{Prelude to Compressed Baryonic Matter}

\author{Frank Wilczek \\
\small\it Center for Theoretical Physics\\[-1ex]
\small \it Department of Physics\\[-1ex]
\small \it Massachusetts Institute of Technology\\[-1ex]
\small\it Cambridge, MA 02139}
\date{}

\maketitle

\begin{abstract}
This is intended to appear as the introduction to ``The CBM Physics Book: compressed baryonic matter in laboratory experiments'' (ed. B. Friman, C. H\"ohne, S. Leupold, J. Knoll, J. Randrup, R. Rapp, P. Senger), to be published by Springer.  At the end there is a new proposal for numerically tractable models of interacting many-body systems.
\end{abstract}

\newpage
\frenchspacing

Why study compressed baryonic matter?  Most obviously, because it's an important piece of Nature.  The whole universe, in the early moments of the big bang, was filled with the stuff.  Today, highly compressed baryonic matter occurs in neutron stars and during crucial moments in the development of supernovae.   Also, working to understand compressed baryonic matter gives us new perspectives on ordinary baryonic matter, i.e. the matter in atomic nuclei.  But perhaps the best answer is a variation on the one George Mallory gave, when asked why he sought to scale Mount Everest: Because, as a prominent feature in the landscape of physics, it's there.    Compressed baryonic matter is a material we can produce in novel, challenging experiments that probe new extremes of temperature and density.   On the theoretical side, it a mathematically well-defined domain with a wealth of novel, challenging problems, as well as wide-ranging connections.    Its challenges have already inspired a lot of very clever work, and revealed some wonderful surprises, as documented in this volume.  

Despite -- or rather because of -- all this recent progress, I think the best is yet to come.   Central questions have not been answered. What is the phase diagram?  What is in the interior of a neutron star?  And more broadly: How can we make better use of QCD in nuclear physics?  Its equations should, in principle, contain all the answers; but in practice we struggle.  With more powerful accelerators and detectors, bigger and faster computers, and improved insight, we are advancing on these issues.   The latest investigations also raise new questions:  Why does quark-gluon plasma, near its crossover temperature, behave remarkably like a perfect liquid, perhaps well described by a simple strong-coupling theory?   Does cold baryonic matter at ultrahigh density really form the color-flavor locked state, where weak coupling methods borrowed from the theory of superconductivity describe the physics of confinement and chiral symmetry breaking?   Can we build out from the profound simplicity apparently discovered in those idealizations, to make them more realistic and comprehensive?  

Part of the appeal of the study of compressed baryonic matter is that it presents four distinct aspects: analytic theory, laboratory experiments, astrophysical observations, and last but certainly not least numerical ``experiments''.   I'd like to make a few comments about each of these aspects, in turn.   

\bigskip

{\it Analytic Theory}

\bigskip

The foundation of QCD is uniquely secure.  Once the numerical values of quark masses and an overall coupling are given, the equations of QCD are defined with mathematical precision\footnote{If neither of the discrete symmetries parity P or time reversal T are imposed, one must also specify the notorious $\theta$ term.  Empirically, this term is known to be very nearly zero.}.     Quantum theory, special relativity, and the profound embodiment of those principles in gauge field theory, impose such powerful consistency constraints that the theory does not allow any significant modification nor additional parameters.  

Indeed, the situation is even a little better than the preceding counting of parameters indicates, due to the phenomenon of dimensional transmutation.  Let me take a moment to elaborate on that profound feature of QCD.  Since we are discussing a theory wherein quantum mechanical fluctuations and relativistic motion are all-pervasive, it is appropriate to adopt Planck's constant $\hbar$ and the speed of light $c$ as our units for action and velocity, so that $\hbar = c =1$. In this system, the unit of any physical quantity is some power of mass, its mass dimension.  With that understanding, we can appreciate the power of the dimensional transmutation principle, which is this: 

{\it Dimensional Transmutation Principle}: If, in the formulation of QCD, one re-scales all the quark masses by a common constant, and at the same time makes an appropriate (nonlinear) change in the coupling constant, one obtains a new theory in which all {\it dimensionless\/} -- that is, numerical -- quantities retain the same values as in the original theory.   

In other words, the new theory differs from the old only in an overall change of the unit of mass, or equivalently length$^{-1}$.  The dimensional transmutation principle is an immediate consequence of the fact that the coupling ``constant'' is not simply a numerical quantity; its definition brings in a length scale.   We could, for instance, define the coupling constant as the ``charge'' appearing in the ``Coulomb's law'' for forces between quarks at short distances; but since in QCD the analogue of Coulomb's law is not accurately valid, we obtain a distance-dependent, running coupling constant.    We can freely specify some conventional value  $g_0$ for  the coupling.  The physically defined coupling $g(l)$, which depends on the distance $l$, takes this value at some length $l_0$, where $g(l_0) = g_0$.   Thus we can trade the conventional, numerical quantity $g_0$ for the physical quantity $l_0(g_0)$ it points to.   But $l_0$ has dimensions of length, or inverse mass.   So what might have appeared to be a freely specifiable numerical quantity is actually another freely specifiable mass parameter, on the same footing as the quark masses.    If we rescale all the masses by a common factor, we have merely changed the standard by which we measure mass (or length).   What might have appeared to be a one-parameter family of essentially different theories, with different values of the coupling constant, is actually just one theory, viewed through lenses of differing magnification.  If we discount that trivial difference, one additional parameter can be disposed of.    

{\it QCD Lite\/} is the version of QCD wherein  the masses of the $u, d,$ and $s$ quarks are taken to zero and the masses of the $c, b$ and $t$ quarks are taken to infinity (so that in effect the heavy quarks are neglected).   We expect on theoretical grounds that QCD Lite provides an excellent qualitative and a reasonable semi-quantitative approximation to real-world QCD in many applications. In particular, QCD Lite supports confinement and chiral symmetry breaking and it yields, for baryon number $|B| \leq 1$, an excellent approximation to the spectrum of real-world QCD\footnote{However the quark-hadron continuity principle \cite{schaeferW}, which is closely connected to color-flavor locking, suggests that ``Nuclear Physics Lite'', for $|B| > 1$, is quite different from real nuclear physics.}.   Those expectations are borne out by numerical studies of the two theories.   

There are a handful of cases where we can make analytic predictions in QCD.   The most basic and most useful, by far, is in the regime of short-distance or high-energy processes, where we can apply renormalization group methods.  We can exploit asymptotic freedom, and calculate short-distance or high-energy behavior using ordinary perturbation theory around free quark and gluon fields \cite{perturbative}.    It is not always obvious (to say the least) what aspects of a complex process involving hadrons really reflect short-distance behavior.   Over the years the scope of perturbative QCD has expanded enormously, from its original base in deep inelastic scattering and other current-induced processes (the Age of Operators) to include many aspects of jet physics, heavy quark physics, and more.   In the modern Age of Quarks and Gluons, the operating principle is the Jesuit Credo, that it is more blessed to ask forgiveness than permission.   One starts by assuming tentatively that the quantity of interest, such as (say) the probability of producing a jet within a given energy range and solid angle, can be calculated by pretending that quarks and gluons are the physical degrees of freedom of QCD, and then checks whether higher orders in perturbation theory render that starting assumption untenable.    This principle has been extremely fruitful and successful, giving us valuable and -- so far -- reliable guidance regarding the applicability of perturbation theory.   These ideas dominate the interpretation of accelerator experiments at the high energy frontier; for example, experimenters routinely speak of measuring cross-sections for producing quarks or gluons, meaning of course the jets calculated using those concepts.   

It's long been known that perturbation theory can also be applied to describe QCD at asymptotically high temperatures, to calculate such things as the energy density and pressure.   Intuitively, keeping in mind the Jesuit Credo, this seems utterly reasonable: At high temperatures, taking noninteracting quarks and gluons as the starting point, the corrections to bulk properties due to interactions seem likely to be small, since most of those quarks and gluons will be highly energetic, and thus weakly interacting. 
Indeed, until fairly recently it was widely anticipated that the temperatures required to produce a nearly ideal gas of quarks and gluons
might not be terribly large.  After all, in deep inelastic scattering approximate free-field behavior sets in at remarkably low energies --  this is the phenomenon of precocious scaling.    The truth is more interesting, as we'll discuss further below.   

A different regime of highly compressed baryonic matter supports a most remarkable analytic theory \cite{alfordRSS}.   This is the regime of large chemical potential, or in other words high baryon number density, and low temperature.   Here methods borrowed from the BCS theory of superconductivity come into play.    Starting from free quarks, one derives large fermi surfaces.  Therefore the relevant low-energy excitations, involving transitions between occupied and unoccupied states near the fermi surface, involve quarks with large momenta.   Such quarks can be expected to be weakly interacting, so we seem -- at first glance -- to have a simple ``quark soup''.   There are two problems with that picture, however.   First, it fails to consider the gluons.   And the gluons are indeed problematic.  While their electric interactions are screened by the quarks, their magnetic interactions are not screened, and the unscreened magnetic interactions among gluons lead to infrared divergences, invalidating our hypothesis of weak coupling.   Second, it fails to take into account that there are many low-energy quark-quark excitations with the same quantum numbers.  This means that we're doing {\it degenerate\/} perturbation theory among those states.   (This too is signaled by infrared divergences.)   In degenerate perturbation theory, even small perturbations can make qualitative changes.    Specifically, quark-quark pairs with equal and opposite 3-momenta $\pm \vec p$ are subject to the same Cooper pairing instability that leads, in ordinary condensed matter systems, to superconductivity in metals or  to superfluidity in liquid He$^3$.   Remarkably, these two problems can be resolved analytically.    The quarks have an attractive interaction, which triggers a version of superconductivity for the color gauge interaction, analogous to ordinary superconductivity for electromagnetism.   In more detail: Quarks form a pairing condensate.  For three quarks with negligible masses a particularly simple and beautiful symmetry breaking pattern, the so-called color-flavor locking, appears to be energetically favored.   In this state the quarks pair as
\begin{eqnarray}
\langle q_{L}^{a\alpha \mu} q_{L}^{b\beta \nu} \rangle \, &=& \, -\langle q_{R}^{a\alpha \mu} q_{R}^{b\beta \nu} \rangle  \\
\, &=& \, \epsilon^{\mu \nu} \bigl( \kappa_{\bar 3} (\delta^{a\alpha} \delta^{b\beta} - \delta^{a\beta} \delta^{b\alpha} ) + \kappa_6  (\delta^{a\alpha} \delta^{b\beta} + \delta^{a\beta} \delta^{b\alpha} ) \bigr) \nonumber
\end{eqnarray}
Here $\mu, \nu$ are Dirac spinor indices, $\alpha, \beta$ are color indices, and $a, b$ flavor indices, $\kappa_{\bar 3} >> \kappa_6$ are coefficients of pairing in antitriplet and sextet color channels, and momentum dependence has been suppressed.   This pairing has many interesting features.  It breaks the color gauge and chiral flavor symmetries down to a diagonal global flavor group, according to
\begin{equation}
SU(3)_c \times SU(3)_L \times SU(3)_R \, \rightarrow \, SU(3)_{\Delta}
\end{equation}
where the residual symmetry transformations are of the form $(g, g^*, g^*)$.  The ``spontaneous breaking'' of color symmetry produces gaps in both the gluon and the quark sector.   The correlation between left-handed ($L$) and right-handed  ($R$) quark fields breaks chiral symmetry.   The modified form of the symmetry generators has as a consequence that the elementary excitations around the new ground state have integer electric charge; and indeed the whole spectrum of low-lying states resembles what one expects for the confined phase of this theory.   So we have here a weak-coupling implementation of the main qualitative features that distinguish perturbative from nonperturbative QCD!   The deep point is that the color-flavor locked ground state is constructed non-perturbatively, following the ideas of the Bardeen-Cooper-Schrieffer (BCS) theory of superconductivity -- here weak interactions, in the context of degenerate perturbation theory, do indeed trigger drastic reorganization of the ground state.

Very recently another front of analytic work on compressed QCD has opened up \cite{adsCFT}.   This is inspired, on the theoretical side, by the AdS/CFT correspondences of string theory.   In the right circumstances, those correspondences allow one to map strong coupling problems in four-dimensional gauge field theory onto problems in five-dimensional classical general relativity.   On the experimental side, we have indications that the initial fireball of material produced in heavy ion collisions, which probably broadly resembles QCD near equilibrium at $T \sim 150-200$ MeV, is characterized by strong effective coupling.  The observed fireball behaves as a near-perfect ideal liquid, rather than an ideal gas!   Unfortunately QCD is not a case where AdS/CFT applies directly, nor is it manifestly close to one, so considerable guesswork is involved in application of  AdS/CFT ideas to reality.   It's still early days in this field, however, and the possibility of connecting five-dimensional general relativity to observations is so startling and deep that it must be pursued vigorously.   

Finally, there is another class of analytic theory that can be applied to the QCD of condensed matter \cite{rajaW}.  This is the theory of second-order critical points.   Both the strength and the weakness of this theory is that it is so general (``universal''); it is largely independent of the microscopic dynamics.     Second-order critical points are associated with change in symmetry, and the low-energy, long-wavelength behavior near such critical points is dominated by critical fluctuations.  Those modes and their interactions are described by effective theories, which depend only on the symmetries (here: chiral symmetry).  The effective theories relevant to second-order transitions in QCD are much simpler than QCD itself, but they can be exploited to make detailed quantitative predictions that should also hold in QCD itself.

\bigskip

{\it Cosmology and Astrophysics}

\bigskip

As I mentioned earlier, compressed baryonic matter filled the early universe, and is the stuff of neutron stars and supernovae.   

QCD and asymptotic freedom, by enabling us to extrapolate and draw consequences from the laws of physics to the ultra-extreme conditions of the early big bang with confidence and precision, provide the intellectual foundation for modern early universe cosmology. At present, however, the link between cosmological or astrophysical observations and fundamental theory is quite tenuous.    In cosmology, I'm afraid this situation appears unlikely to change\footnote{I'd love to be wrong about that!}.   The problem is that the evolution of the universe, even in its dynamic early stages, is very slow on strong-interaction time scales, so that accurate equilibrium is likely to maintained.  If at any stage the equilibrium were inhomogeneous, due to phase separation at a first-order transition or even fluctuations at a second-order transition, then one might hope that some historical memory would be imprinted on nuclear abundances, or in gravity waves.   But the emerging consensus is that real-world QCD at very small chemical potential, which is what we encounter in conventional cosmology, does not support either sharp phase transitions or large equilibrium fluctuations.    So, barring a very major deviation from orthodox expectations (e.g., Affleck-Dine baryogenesis \cite{affleckD}) no observable relic of the early strong dynamics seems likely to survive.  

Astrophysics is, I think, much more promising.    It's not entirely ridiculous to think that in the foreseeable future we'll have mass-radius relations for a good sample of neutron stars, rich measurements of neutrino emission in a few supernova explosion, and measurements of gravity wave emission accompanying the final infall of binary systems that include neutron stars.   These signals, and perhaps others, will in coming years provide a rich flow of information from the world of compressed baryonic matter.   We should get ready for it!

There's also an ongoing revolution in the experimental study of high-energy cosmic rays.  Perhaps there are opportunities for extreme QCD here, either in the description of the interactions of heavy nucleus primaries or in modeling the sources; but these are complex subjects, where QCD contributes to only a small share of the uncertainties.

\bigskip

{\it Laboratory Experiments}

\bigskip

I shall be very brief here, since the material on this topic in the body of the book speaks ably for itself.    

Although by now we almost take it for granted, it's important to remember that the most fundamental prediction of QCD at high temperatures is both simple and spectacular: 

At low temperatures temperatures strongly interacting matter should be well described using separate hadrons.  In fact  below $T \lesssim 125$ MeV or so, it should be basically a gas of pions, with 3 bosonic degrees of freedom.  But at high temperatures, strongly interacting matter should be described as a plasma of quarks and gluons, with with 36 fermionic and 16 bosonic degrees of freedom!  (Three flavors and three colors of quarks and antiquarks, each with two spin states; eight gluons each with two polarization states.)   In fact we find, both upon numerical solution of the theory and in experiments, that within a remarkably narrow range of temperatures things do change dramatically.  By $T \gtrsim 170$ MeV or so the energy density is pretty nearly that of a gas of free quarks and gluons, with 36 fermionic and 16 bosonic degrees of freedom!   This enormous quantitative change, of course, serves to dramatize the qualitative change that underlies it. Quarks and gluons, once famous for being confined and elusive, have dropped their masks and come to center stage.  

While this zeroth-order success is gratifying, closer examination reveals that the situation is far from straightforward.   It's long been known, from the numerical work, that while the energy density quickly approaches its predicted asymptotic value, the pressure lags.   So the weak-coupling theory is suspect near the transition temperature.   (Though tour-de-force calculations in perturbation theory, involving high orders and some resummations, work significantly better.)   And here experiment produced perhaps its greatest surprise: while the {\it thermodynamic\/} properties of the plasma near $T \sim 150$ MeV are not so different from those of an ideal gas of quarks and gluons, the {\it transport\/} properties point to very strong interactions.   Indeed, the plasma appears to have very small shear viscosity, so that it more resembles an ideal liquid.   Were to think in terms of standard kinetic theory -- which is no doubt too naive -- this phenomenon would point to a tiny mean free path.   

Clearly, it will be interesting to observe whether, or to what extent, these surprising properties survive at the higher temperatures that will be explored in heavy ion collisions at the LHC.   Speaking more broadly, it will be highly interesting to explore how the asymptotically free ultra-short distance and ultra-high energy behavior of hadronic matter connects to the distinctly different intermediate-scale and overall transport behavior.   At LHC there will be enough phase space and luminosity to exploit $c$ and even $b$ quarks, and very high energy photons, dimuons, and jets, with rich statistics, and really engage these issues.

Another important goal of future experiments is to map out the phase structure of hadronic matter, as a function of temperature {\it and\/} baryon number density.    In this regard an interesting target is a possible true second-order phase transition (a tricritical point).   Such a transition should be accompanied by critical fluctuations, which might leave signatures  visible even within the daunting environment around heavy ion fireballs.    These possibilities will be explored both at the Relativistic Heavy Ion Collider (RHIC) and the new Facility for Antiproton and Ion Research (FAIR).   

Finally, its appropriate to mention that the classic experimental probe of hadronic structure, deep inelastic scattering, remains interesting and relevant.   The ``gluonization'' of hadrons, as they are analyzed by virtual photons of large $Q^2$, is a profound phenomenon, that connects in important but poorly understood ways to dominant interaction mechanisms between high-energy hadrons (diffractive scattering and diffractive dissociation).   These issues can also be studied for nuclei, of course, and should give insight into some fundamental questions.  Is there a dynamically defined variety of quark, in a piece of the nuclear wave function, that is shared among different nucleons, or are all of them, uniformly, strictly confined?    I sense that this a domain where accurate experimental work might connect with new analytical ideas.

\bigskip

{\it Numerical Experiments}

\bigskip

Numerical solution of the equations of QCD, usually called lattice gauge theory, is a subject that has achieved major triumphs \cite{latticeText}.    Confinement and spontaneous chiral symmetry breaking have been demonstrated convincingly.  Going far beyond those qualitative results, it has been demonstrated by direct calculation that the rigid microscopic equations of QCD, based on quarks and gluons, reproduces the rich spectrum of low-energy hadrons that is observed in Nature.  Numerical work also revealed the striking temperature dependence of the energy density and pressure, mentioned earlier, well in advance of its experimental confirmation \cite{latticeT}.  No other technique on the horizon comes close to competing with the use of sophisticated discretization techniques, clever algorithms and powerful computers for addressing these and many other issues (e.g., evaluating matrix elements of weak and electromagnetic currents, heavy quark spectroscopy) quantitatively.   
Where lattice gauge theory can be applied, it is generally unrivaled.  

Unfortunately, however, lattice gauge theory is not well adapted to deal with some other central questions and opportunities in QCD.    All known algorithms degrade badly for problems that cannot somehow be phrased as the evaluation of positive-definite integrals -- in gigantic spaces, to be sure.    But the partition function of QCD at finite chemical potential is, when cast into an integral, not positive definite; one has the notorious fermion sign (actually phase) problem.   The final answer is much smaller than the various, largely canceling, contributions to it.   Existing numerical techniques thereby lose most of their power, and today we have no meaningful ability to investigate QCD at high baryon number density by direct solution of equations, even though we know exactly what the equations are.   

Different but equally severe problems appear when we try to investigate high-mass, high-spin resonances.   That is a pity, because QCD exhibits remarkably simple regularities in the high-spin regime, so it's tempting to conjecture that a good analytic or semi-analytic theory awaits discovery.  

I'd like to take this opportunity to mention a pet idea of mine, that I think merits attention.   As just mentioned, known numerical techniques in quantum field theory are extremely powerful when applied to positive-definite integrals, though they founder for oscillatory ones.    This difficulty creates a barrier to direct simulation of nuclear matter or investigation of its phase diagram.   There is, however, the possibility of dodging it, by pursuing what I call ``lattice lattice gauge theory''.   That is, one can insert sources belonging to  {\it real\/} representation of the gauge group.  These do not lead to sign or phase problems.   Thus lattices of fixed sources in real representations provide user-friendly model systems.   The spacing of the source lattice can and should be varied independently of the spacing in the numerical grid; this is what I mean by lattice lattice gauge theory.    We can vary the dimensionality, the type of lattice, the gauge group(s), and the representation(s) present;  we can also include temperature or even (at a high price) dynamical fermions.   

As a concrete example, we could consider putting octet sources on a cubic lattice in pure glue QCD.   At large separation these sources will induce a lattice of individuated glueballs.  At small separation, the glueballs will start to overlap, and presumably there will be a transition to a screening ``metallic'' state.   This is an analogue of the Mott transition, which is a central research area in today's condensed matter physics.   There might also be an intermediate state with flux directed between pairs of neighboring sources, in the spirit of Anderson's resonating valence bond (RVB) hypothesis.   The point is we don't have to guess, we can calculate.   

In QCD models of this kind could shed light on the nature of glueballs or (with dynamical fermions) $Gq\bar q$.  They could also supply a {\it very\/} crude, but tractable, caricature of nuclear matter, and it would be interesting to compare their Mott transition density to the density of nuclear matter.   For exploring quasi-chemical questions, which might be the most fascinating, it would probably be advisable to make do with finite groups, and to invest more heavily in some of the other bells and whistles I mentioned before.   

Even were this program to prove wildly successful, it would of course only be a stopgap.    To get quantitative results for highly condensed baryonic matter, worthy of the challenges posed by Nature and the opportunities afforded by QCD, we will need fundamentally different numerical methods.   In conventional lattice gauge theory the starting point is the no-particle, Fock vacuum.  Probably, as suggested by color-flavor locking, the key is to start with a better approximation to the true ground state, generated numerically through an appropriate variational principle.

\bigskip

\bigskip

As evidenced by this volume, the study of compressed baryonic matter is paying off handsomely.   Old questions are being answered; but the answers lead us to better, more ambitious questions, and leave us hungry for more.

\end{document}